\documentclass[twocolumn,amsmath,amssymb,aps,pra]{revtex4-2}
\usepackage{bm}
\usepackage{amsmath,amssymb}
\usepackage{graphicx}
\usepackage{float}
\usepackage{mathtools}
\usepackage{dcolumn}
\usepackage{hyperref}
\begin{document} 
\title{Constructing various paraxial beams out of regular and modified Bessel-Gaussian modes}
\author{Tomasz Rado\.zycki}
\email{t.radozycki@uksw.edu.pl}
\affiliation{Faculty of Mathematics and Natural Sciences, College of Sciences, Institute of Physical Sciences, Cardinal Stefan Wyszy\'nski University, W\'oycickiego 1/3, 01-938 Warsaw, Poland} 

\begin{abstract}
Various superpositions of Bessel-Gaussian beams and modified Bessel Gaussian beams are considered. Two selected parameters characterizing these beams, with respect to which the superpositions are constructed, are the topological index $n$ associated with the orbital angular momentum carried by the beam, and $\chi$ related to the dilation of the beam. It is shown that, from these modes, by choosing appropriate weighting factors, it is possible to create a number of well- and less-known solutions of the paraxial equation: Gaussian (shifted and non-shifted) beam, $\gamma$ beam, Kummer-Gaussian beam, special hyperbolic Bessel-Gaussian beam, a certain special Laguerre-Gaussian beam, and generalized paraxial beams in hyperbolic and regular versions. 
\end{abstract}

\maketitle

\section{Introduction}\label{intr}

Since when the first solution of the Helmholtz paraxial equation in the form of the Gaussian beam~\cite{kl,davis,sie,nemo,mw,saleh,ibbz,sesh,gustavo,er,selina} was reported, a great variety of other beams with more complicated structures have been analytically derived and then experimentally generated. Although the Gaussian beam (i.e the so called fundamental mode) in many situations accurately describes the laser field that exhibits cylindrical symmetry, many practical applications require somewhat more complex patterns. Enough to mention here other cylindrical beams as regular Bessel-Gaussian (BG) beams~\cite{saleh,she,gori,april1,mendoza} or modified ones (mBG)~\cite{bagini} and Laguerre-Gaussian (LG) beams~\cite{sie,saleh,mendoza,lg,lg2,april2,april3,nas} of various orders which exhibit ring-like structure, Kummer-Gaussian (KG) (i.e.,  Hypergeometric-Gaussian) beams~\cite{kot,karimi} or non-cylindrical beams like for instance Hermite-Gaussian ones~\cite{kl,sie} or more general paraxial beam~\cite{trgen}. Due to their numerous applications ranging from pure physics through optical communication technologies, image processing,  up to biology and medicine~\cite{ste,fazal,pad,woe,bowpa,grier1,kol,alt,nis,cc}, the structured light has earned an extensive literature cited above only in a very nutshell.

The plethora of miscellaneous beams requires a kind of of ordering and establishing some relationships between them. Some efforts to unify the derivation and the description of the paraxial beams have been made in the past~\cite{end,li,vl,fe}. In our previous work~\cite{trhan}, we proceeded along these lines, attempting to exploit the Hankel transformation~\cite{patra,erde} for this purpose. The present work is in a sense devoted to similar issue, albeit from a different point of view.

From the mathematical perspective all of the mentioned modes constitute certain exact solutions of the Helmholtz paraxial equation which by itself is already approximated, but perfectly describing laser beams near the optical axis~\cite{sie,lax}:
\begin{equation}\label{paraxiala}
\mathcal{4}_\perp \psi(\bm{r},z)+2ik\partial_z \psi({\bm{r}},z)=0,
\end{equation}
The scalar function $\psi(\bm{r},z)$ proportional to the field strength, we focus on in this paper, is called the {\em envelope}. The Laplace operator $\mathcal{4}_\perp$ appearing in (\ref{paraxiala}) is the two-dimensional one acting in the transverse plane only (i.e., $\bm{r}=[x,y]$) and the symbol $\partial_z$ stands for the partial derivative $\frac{\partial}{\partial z}$ (and similarly for other variables). 

The two solutions to this equation that will serve as the basic ones in this work are mBG and BG modes. They play an important role due to their almost nondiffractive character and numerous applications, for instance for harmonic generation in nonlinear optics and others. The expressions describing these modes are well known and are given by formulas~(\ref{mbgbeam}) and~(\ref{bgbeam}). 

The BG and mBG beams have one advantage (from the point of view of theoretical description, but also practical applications) over Gaussian beams: it is the occurrence, apart from the orbital angular momentum index $n$, of an additional parameter $\chi$, which can provide an extra basis for the formation of a certain new superpositions. Its appearance in the case of the BG beam is a consequence of its design out of Gaussian beams, as discussed below, in which the latter are characterized by the double opening angle: that associated with the diffraction of constitutive Gaussian beams and also their angle of inclination with respect to the $z$-axis. In the case of the mBG beams, the role of this latter angle is played by the radius of the cylinder on which the wave vectors are distributed.

When dealing with BG and mBG beams in the further part of the paper, we ignore normalization factors. While this does not affect the conclusions drawn from the calculations, it greatly simplifies the emerging expressions. In general, the normalization factors ($N$) are rather complicated functions of the parameters $\chi$ and $n$ (among others). Including them in superpositions would merely lead to redefinition and unnecessary complication of the weighting factors ($w\mapsto w/N$), without affecting any meaningful conclusions. The same remark also applies to other beams occurring in this work. Consequently, any comments on the behavior of expressions describing a given beam for large values of $\chi$ or $n$ refer, therefore, to {\em unnormalized} expressions explicitly stated in the work. The experimental realization of the interference of waves requires a separate adjustment of their relative intensities anyway.

The next two sections present a series of superpositions leading to other beams known from the literature. We start, in Section~\ref{supmbg} with providing superpositions of mBG beams, owing to the fact that they exhibit better convergence properties as $\chi\rightarrow\infty$, than BG beams do, due to the presence of the factor $\exp[-\chi^2/\alpha(z)]$. For the latter beams, dealt with in Section~\ref{bgb}, the sign of this exponent is inverted, which necessitates the inclusion of an additional Gaussian damping function in the weighing factors and consequently precludes superpositions in which these factors are purely power-law in nature. For the two most transparent cases, the effect of superposition is visualized in the figures.

\section{Superpositions of modified Bessel-Gaussian beams}\label{supmbg}

The modified Bessel-Gaussian beam of the $n$th order has the following form:
\begin{equation}\label{mbgbeam}
\Psi_{mBG}^{(n)}(r,\varphi,z)=\frac{2}{\alpha(z)}e^{in\varphi}e^{{\textstyle -\frac{\chi^2+r^2}{\alpha(z)}}}I_n\left(\frac{2\chi r}{\alpha(z)}\right),
\end{equation}
where, as mentioned in the Introduction, the normalization factor has been omitted. Here 
\begin{equation}\label{alpha}
\alpha(z)=w_0^2+2i\,\frac{z}{k}=w_0^2\left(1+i\frac{z}{z_R}\right),
\end{equation}
is the complex beam parameter, $r$ stands for $\sqrt{x^2+y^2}$, $w_0$ denotes the beam waist (i.e. the distance from the propagation axis at which the irradiance weakens $e^2$ times), which is specified by the Gaussian factor, and $z_R=kw_0^2/2$ is the Rayleigh length, (i.e. the distance at which the perpendicular ``area'' -- $\pi w_0^2$ -- related to this Gaussian factor duplicates). These quantities recur below in all beams containing a Gaussian factor, although their meaning may vary slightly from case to case.

The mBG beam can be constructed as equal-weight superpositions of Gaussian beams~(\ref{gauss}) but with foci located on a circle of radius $\chi$~\cite{bagini}. The parameter $\chi$ is then arbitrary and remains at our disposal (formally from zero to infinity, but in realistic experimental situation there must occur some truncation) in the sense that expression (\ref{mbgbeam}) constitutes the solution of the paraxial equation regardless of its value. This property allows for constructing further superpositions of modes with differing values of $\chi$ and with weighting factors that can be $\chi$-dependent. The other kind of superpositions is obtained by interfering modes with equal $\chi$ and varying values of the topological index $n$.

\subsection{Gaussian beam}\label{gaumbg}

The $n$th order Gaussian beam has the well-known form:
\begin{equation}\label{gauss}
\Psi_G^{(n)}(r,\varphi,z)=\frac{1}{\alpha(z)^{n+1}}\,r^ne^{in\varphi}e^{{\textstyle -\frac{r^2}{\alpha(z)}}},
\end{equation}
where $\alpha(z)$ is defined in~(\ref{alpha}).
In this subsection it is first demonstrated that the sum of the co-focal mBG modes~(\ref{mbgbeam}) with appropriate $\chi$-dependent weights restores the Gaussian beam, in general with focus shifted along the $z$ axis and/or of modified waist value. 

Let us consider the following superposition obtained by integrating over the parameter $\chi$ with relative amplitudes chosen in the form given below:
\begin{equation}\label{supgmbg}
\Psi_s(r,\varphi,z)\stackrel{\mathrm{def}}{=}\int\limits_0^\infty d\chi\left(\frac{\chi}{\kappa}\right)^{n+1}\!\!\!\! e^{{\textstyle -\frac{\chi^2}{\kappa}}}\Psi_{mBG}^{(n)}(r,\varphi,z).
\end{equation}
The quantity $\kappa$ is a certain constant (satisfying the condition $w_0^2+\mathrm{Re}\,\kappa>0$), the interpretation of which will be given later, and the subscript $s$, here and below, stands for ``superimposed''.

Apparently this expression seems to be singular as $\kappa\rightarrow 0$, but the close examination shows that this limit is well defined, which is owed to the behavior of the Bessel function $I_n$ close to the origin. This is confirmed by the explicit calculation below.

\begin{figure*}[!]
\begin{center}
\includegraphics[width=0.9\textwidth,angle=0]{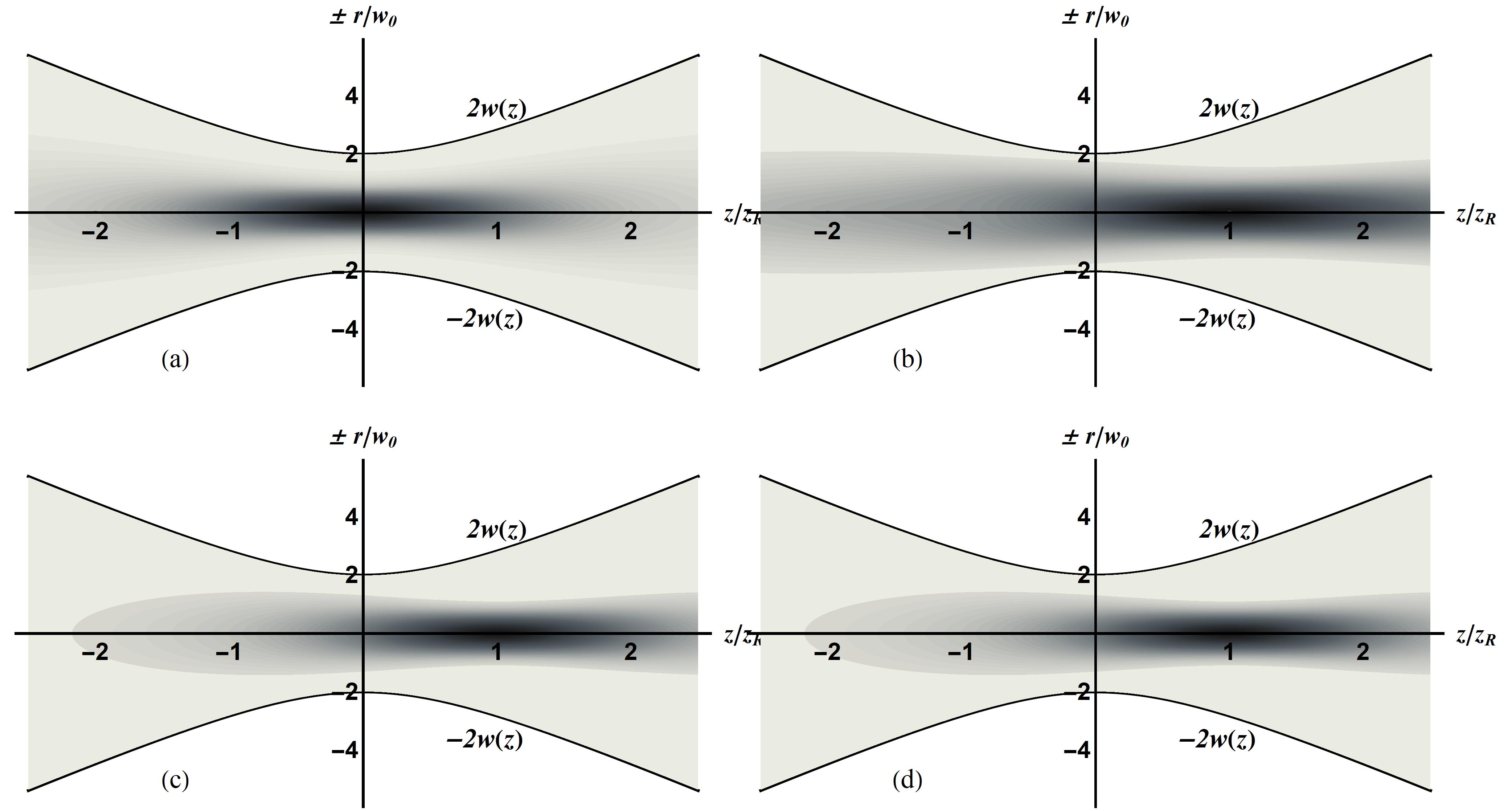}
\end{center}
\vspace{-4.5ex}
\caption{Intensity of superimposed mBG beams, according to the formula~(\ref{supgmbg}) for $n=0$ and $\kappa=-2iz_R/k$ in the cut plane containing the propagation axis. Subsequent plots are performed for increasing number of superimposed modes: (a) one mode with $\chi= 0.01z_R$, (b) $10$ modes with $\chi= 0.01z_R, 0.02z_R,\ldots,0.1 z_R$, (c) $100$ modes with $\chi= 0.01z_R, 0.02z_R,\ldots,1 z_R$, (d) $2000$ modes with $\chi= 0.01z_R, 0.02z_R,\ldots,20 z_R$. The images are drawn in negative form for better visibility: black color corresponds to high, and white to low intensity. Auxiliary lines $\pm 2w(z)$ are drawn as well.}
\label{cmbgfig}
\end{figure*}

Substituting expression~(\ref{mbgbeam}) into~(\ref{supgmbg}), and using the well-known integral, convergent for $\mathrm{Re}\,a>0$, which takes the simple form for natural $n$~\cite{prudnikov}:
\begin{equation}\label{int1}
\int\limits_0^\infty dx\, x^{n+1}e^{- a x^2} I_n(b x)= \frac{b^n}{(2a)^{n+1}}\,e^{{\textstyle \frac{b^2}{4a}}},
\end{equation}
where it can be set
\begin{equation}\label{ab}
a=\frac{1}{\kappa}+\frac{1}{\alpha(z)},\;\;\;\;\;\; b=\frac{2r}{\alpha(z)},
\end{equation}
one obtains
\begin{equation}\label{sgmbg}
\Psi_s(r,\varphi,z)=\frac{1}{(\alpha(z)+\kappa)^{n+1}}\,r^ne^{in\varphi}e^{{\textstyle -\frac{r^2}{\alpha(z)+\kappa}}}
\end{equation}
This expression represents the standard form of the $n$th order Gaussian beam~(\ref{gauss}) with the complex parameter $\alpha(z)$ shifted by $\kappa$:
\begin{equation}\label{ashiftp}
\alpha(z)\;\;\longmapsto\;\; \alpha(z)+\kappa.
\end{equation}
Referring to the formula~(\ref{alpha}), it is clear that the real part of $\kappa$ merely modifies the beam waist, and imaginary part is responsible for the shift of the beam along the $z$ axis:
\begin{equation}\label{mobw}
w_0^2\;\;\longmapsto \;\; w_0^2+\mathrm{Re}\,\kappa,\;\;\;\;\;\; z\;\;\longmapsto \;\; z+\frac{1}{2}\,k\,\mathrm{Im}\,\kappa,
\end{equation}
i.e. the new focus is located in the plane $z=-\frac{1}{2}\,k\,\mathrm{Im}\, \kappa$.
These results are understood on the ground of geometrical and physical considerations. Due to the presence of the factor $\exp[-\chi^2(\frac{1}{\kappa}+\frac{1}{\alpha(z)})]$ under the integral, the values of the expression in parentheses that are significantly distinct from zero give a substantial contribution to the integration only in the region of small $\chi$, either owing to the Gaussian nature of the integrand function (real part) or due to its fast oscillations (imaginary part) which entail for the destructive interference. Tiny values of $\chi$, according to the geometric interpretation provided earlier, correspond to beams focused close to the $z$-axis. The maximum of the irradiance should, therefore, at least for $n=0$, occur at a spot located on the $z$-axis for which $z=-\frac{1}{2}\,k\,\mathrm{Im}\, \kappa$ (if $n\neq 0$ the factor $r^n$ in~(\ref{gauss}) modifies the behavior close to the $z$-axis so as the ring of maximal intensity around this point is created), characterized by the new waist size of $\sqrt{w_0^2+\mathrm{Re}\,\kappa}$.

\begin{figure*}[!]
\begin{center}
\includegraphics[width=1\textwidth,angle=0]{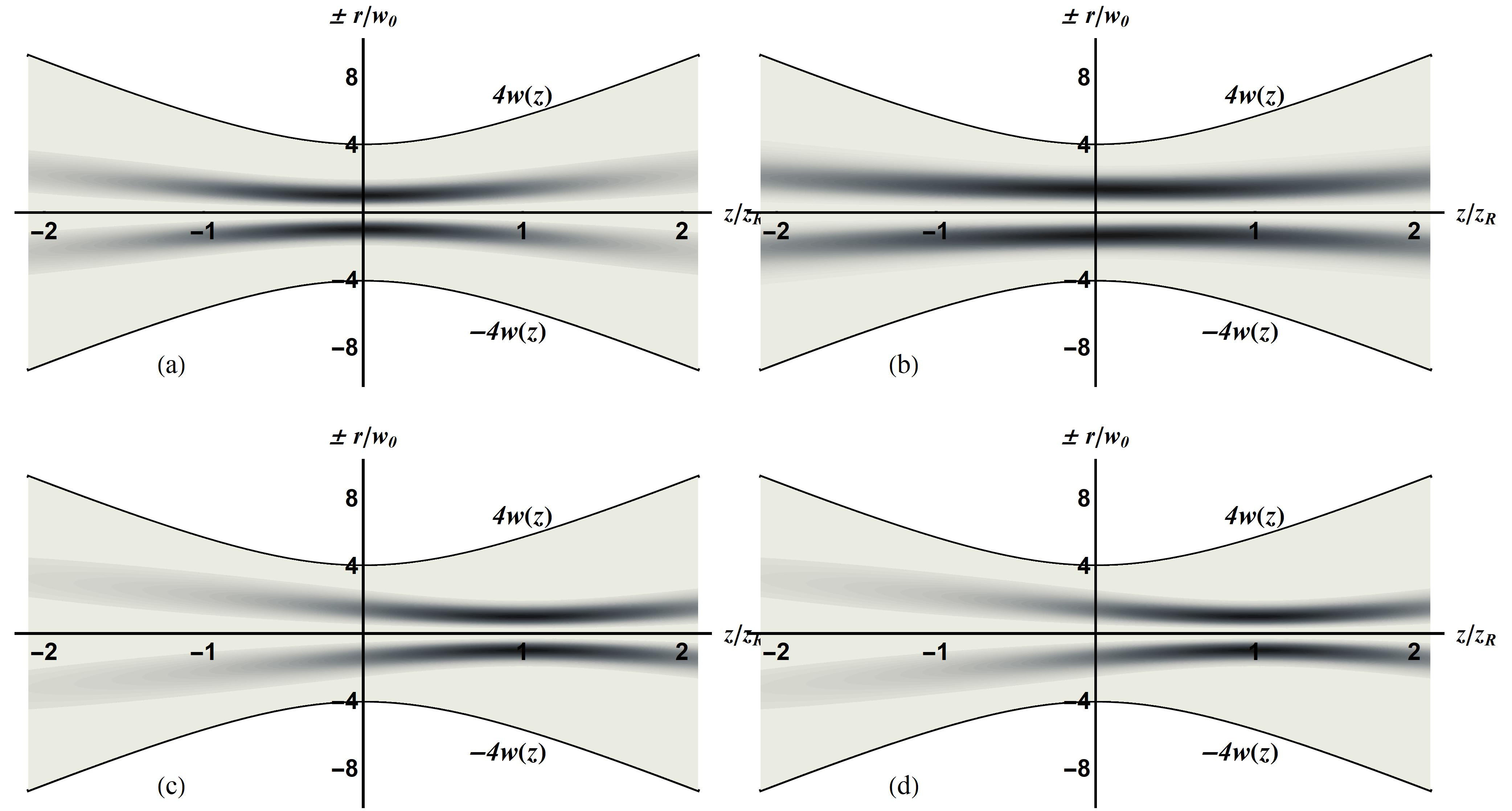}
\end{center}
\vspace{-4.5ex}
\caption{Same as Fig.~\ref{cmbgfig} but for $n=2$ and auxiliary lines corresponding to $\pm 4w(z)$.}
\label{mbgfig}
\end{figure*}

These effects are demonstrated in Figures~\ref{cmbgfig} and~\ref{mbgfig} for $n=0$ and $n=2$ respectively. The superpositions of $1$, $10$, $100$ and $2000$ modes with weighting factors taken from the formula~(\ref{supgmbg}), and uniformly distributed with respect to $\chi$ (so as to avoid modifying these weighing factors) are created. The visible effect is the appearance of the irradiance maximum close to $z=z_R$. This remains in line with our formulas, since the parameter $\kappa$ is chosen in figures to be equal to $-2iz_R/k$, which, according to~(\ref{mobw}), leads to the shift of the Gaussian-beam focus along the propagation axis to $z=z_R$.

Let us now consider the superposition of co-focal mBG modes with fixed $\chi$ but bearing different topological indices $n$ and weighted by factors in the form of $e^{n v}$, with $v$ being a parameter. Thus, let us define
\begin{equation}\label{supgmbgn}
\Psi_s(r,\varphi,z)\stackrel{\mathrm{def}}{=}\sum_{n=-\infty}^\infty e^{nv}\Psi_{mBG}^{(n)}(r,\varphi,z).
\end{equation}
Using~(\ref{mbgbeam}) this sum can be expanded as
\begin{equation}\label{supgnd}
\Psi_s(r,\varphi,z)=\frac{2}{\alpha(z)}e^{{\textstyle -\frac{\chi^2+r^2}{\alpha(z)}}}\sum_{n=-\infty}^\infty e^{n(v+i\varphi)}I_n\left({\textstyle\frac{2\chi r}{\alpha(z)}}\right).
\end{equation}
The well-known expression for the generating function for modified Bessel functions ~\cite{abra}
\begin{equation}\label{generf}
e^{{\textstyle\frac{z}{2}}\left(t+{\textstyle\frac{1}{t}}\right)}=\sum_{n=-\infty}^\infty I_n(z)t^n
\end{equation}
allows us to rewrite~(\ref{supgnd}) in the compact form
\begin{eqnarray}
\Psi_s(r,\varphi,z)=&&\frac{2}{\alpha(z)}e^{{\textstyle-\frac{\chi^2+r^2}{\alpha(z)}}}\label{supgnc}\\
&&\times e^{{\textstyle\frac{2\chi r}{\alpha(z)}}\left(\cos\varphi\cosh v+i\sin\varphi\sinh v\right)}\nonumber\\
&&=\frac{2}{\alpha(z)} e^{{\textstyle-\frac{1}{\alpha(z)}}\,\left[(x-\chi\cosh v)^2+(y-i\chi\sinh v)^2\right]}\nonumber\\
&&=\Psi_{G\;\mathrm{sh}}^{(0)}(r,\varphi,z)\nonumber.
\end{eqnarray}
The obtained expression is known to represent the so called {\em shifted} Gaussian beam~\cite{ji} (this is what `sh' stands for) of the $0$th order. For real values of $v$ the shift in the $x$ direction is real, and that in the $y$ direction -- imaginary. In this derivation, however, nothing prevents us from using complex values of $v$. A purely imaginary value, i.e., $v=i\beta$, for certain real angle $\beta$, shifts the focus radially by the vector $\chi[\cos\beta,\sin\beta]$. Complex shifts in general lead to slanted beams in some aspects similar to that of Fig.~2 of~\cite{kou}.

Thus~(\ref{supgmbgn}) and~(\ref{supgnc}) entail that the shifted Gaussian $0$th-order beam can be expressed as the superposition of co-focused mBG modes with weights given in terms of the exponential function $e^{nv}$. In order to clarify the underlying mechanism, suppose for definiteness that $v$ is real and positive and for simplicity consider the plane $z=0$. According to the formula~(\ref{supgnc}), the peak of irradiance is shifted from the origin of the system along the $x$-axis toward positive values. This effect can be understood when considering the model function
\begin{equation}\label{mofu}
f_n(r,\varphi)=e^{n v}e^{in\varphi}e^{-a r^2}I_n(b r),
\end{equation}
where $a$ and $b$ are certain positive constants. It is obvious that the expression $e^{-a r^2}I_n(b r)$ exhibits a maximum for some positive value of $r$. The requirement of the derivative to vanish at this point yields the equation 
\begin{equation}\label{deva}
\frac{I_n'(br)}{I_n(br)}=\frac{2a}{b}\, r.
\end{equation}
With increasing $n$ the left-hand side increases as well, which can be easily seen if one uses two well-known identities~\cite{gr}
\begin{subequations}\label{twoi}
\begin{align}
&I_n'(z)=\frac{1}{2}\left(I_{n+1}(z)+I_{n-1}(z)\right),\label{twoia}\\
&2n I_n(z)=z\left(I_{n-1}(z)-I_{n+1}(z)\right).\label{twoib}
\end{align}
\end{subequations}
They allow to rewrite~(\ref{deva}) in the form
\begin{equation}\label{}
\frac{I_{n+1}(br)}{I_n(br)}+\frac{n}{b\,r}=\frac{2a}{b}\, r,
\end{equation}
and, due to the positivity of the modified Bessel functions of the first kind (for positive arguments), entails the inequality
\begin{equation}\label{ine1}
 r^2>\frac{n}{2a}.
\end{equation}
At least starting from certain value of $n$ the location of the maximum increases with $n$. 

Now comparing the left ($x<0$) and the right ($x>0$) half-planes, one observes the appearance of the destructive interference for the former (in particular the factor $e^{in\varphi}$ in~(\ref{mofu}) equals $(-1)^n$ on the negative $x$-semi-axis) and the constructive one for the latter. The irradiance maximum is then shifted to the right. For larger values of $n$ the maxima of the subsequent terms in the sum~(\ref{supgmbgn}) move further and further to the right, as we have established above, and participate with substantially greater weighting factors $e^{n v}$. Consequently, larger values of the parameter $v$ lead to an increasingly distant location of the irradiance peak, which is expressed in the formula~(\ref{supgnc}) by the presence of the function $\cosh v$.

\subsection{$\gamma$ beam}\label{gammbg}

The weighting factor chosen under the integral~(\ref{supgmbg}) is not unique and stays at our disposal provided the convergence of this integral with respect to $\chi$ is ensured. Let us then set it in the form of $1/\chi^{n-1}$, and again superimpose the co-focal mBG beams, additionally assuming that $n\geq 2$, i.e,
\begin{equation}\label{supgammbg}
\Psi_s(r,\varphi,z)\stackrel{\mathrm{def}}{=}\int\limits_0^\infty d\chi\,\frac{1}{\chi^{n-1}} \,\Psi_{mBG}^{(n)}(r,\varphi,z).
\end{equation}
After plugging in the expression for $\Psi_{mBG}^{(n)}$, the above integral can be executed with the use of the known formula~\cite{prudnikov}
\begin{equation}\label{int2}
\int\limits_0^\infty dx\, \frac{1}{x^{n-1}}\,e^{- a x^2} I_n(b x)= \frac{(2a)^{n-1}}{(n-1)!\,b^n}\,e^{{\textstyle \frac{b^2}{4a}}}\gamma\left(n,{\textstyle\frac{b^2}{4a}}\right),
\end{equation}
where $\mathrm{Re}\,a>0$ and $\gamma(n,z)$ stands for the incomplete $\gamma$ function~\cite{abra}. In the following $\mathrm{Re}\,a>0$ is assumed. If the normalizability in the transverse plane is to be maintained, the condition $n\geq 2$ has to be imposed due to the asymptotic behavior of the $\gamma$ function:
\begin{equation}
\gamma(n,w)\sim \gamma_{as}(n,w):=\Gamma(n)-w^{n-1} e^{-w},
\label{asg}
\end{equation}
where $as$ indicates $asymptotic$.  

Plugging the expression for $\Psi_{mBG}^{(n)}(r,\varphi,z)$ into~(\ref{supgammbg}) and using~(\ref{int2}), one obtains the result in the form
\begin{equation}\label{sgammbgn}
\Psi_s(r,\varphi,z)=\frac{1}{(n-1)!\, r^n}\,e^{in\varphi}\gamma\left(n,{\textstyle\frac{r^2}{\alpha(z)}}\right).
\end{equation}
Apart from the inessential multiplicative constant this expression represents the so-called $\gamma$ beam~\cite{trgam}. 

It should be mentioned that the $\gamma$ beam does not exhibit the Gaussian behavior but rather the power-law decay $1/r^n$ (for very large $r$). This is a consequence of the circumstance that in this case the superposed modes are those whose irradiance maximum is distributed over a ring with the radius growing with $\chi$. The exemplary integral for $z=0$ may be rewritten as
\begin{eqnarray}
\int\limits_0^\infty d\chi\,&&\frac{1}{\chi^{n-1}}\,I_n\left({\textstyle\frac{2\chi r}{w_0^2}}\right)e^{\textstyle-\frac{\chi^2+r^2}{w_0^2}}\label{exi}\\
&&=\int\limits_0^\infty d\chi\,\frac{1}{\chi^{n-1}}\left[e^{\textstyle-\frac{2\chi r}{w_0^2}}I_n\left({\textstyle\frac{2\chi r}{w_0^2}}\right)\right]e^{\textstyle-\frac{(r-\chi)^2}{w_0^2}}.\nonumber
\end{eqnarray}
The presence of the Gaussian factor ensures that for a given $r$ the significant contribution to the integral only comes from modes where $\chi\approx r$ with spread of several $w_0$. For large $r$, which we are concerned with, this implies $2\chi r\gg w_0^2$ and the expression in square brackets can be approximated by $w_0/(2\pi r)$~\cite{gr}. Consequently, the value of the integral is proportional to $1/r^{n-1}\cdot 1/r=1/r^n$. This clarifies the interference mechanism between various mBG beams (i.e., exhibiting Gaussian fall-off) which finally leads to the power-law decline.

\subsection{Kummer-Gaussian beam}\label{kgaumbg}

Somewhat more general integral~\cite{prudnikov}
\begin{eqnarray}
&&\int\limits_0^\infty dx\, \frac{1}{x^l}\,e^{-ax^2} I_n(b x)\label{int3}\\
&&\;\;\;\;\;\; =\frac{\Gamma\left({\textstyle\frac{n-l+1}{2}}\right)b^n}{n!\;2^{n+1}a^{(n-l+1)/2}}\;
{}_1F_1\left({\textstyle\frac{n-l+1}{2}},n+1,{\textstyle\frac{b^2}{4a}}\right),\nonumber
\end{eqnarray}
where $n-l>-1$, may be applied to derive the superposition of mBG modes producing the so called 
Kummer-Gaussian beam introduced in~\cite{trhan}: 
\begin{eqnarray}
&&\Psi_s(r,\varphi,z)\stackrel{\mathrm{def}}{=}\int\limits_0^\infty d\chi\,\frac{1}{\chi^{l}} \,\Psi_{mBG}^{(n)}(r,\varphi,z)=\frac{\Gamma\left({\textstyle \frac{n-l+1}{2}}\right)}{n!}\label{supkgmbg}\\
&&\;\;\;\;\times\frac{1}{\alpha(z)^{\textstyle \frac{n+l+1}{2}}}\,r^ne^{in\varphi}{}_1F_1\left({\textstyle\frac{n-l+1}{2}},n+1,{\textstyle\frac{r^2}{\alpha(z)}}\right)e^{\textstyle -\frac{r^2}{\alpha(z)}}.\nonumber
\end{eqnarray} 

The interference mechanism is similar to that leading to the $\gamma$ beam.

\subsection{Special hyperbolic Bessel-Gaussian beam}\label{spec2}

Yet another integral, that can be used, has the following form~\cite{prudnikov}
\begin{equation}\label{int4}
\int\limits_0^\infty dx\, e^{-ax^2} I_n(b x)= \sqrt{\frac{\pi}{4a}}\,I_{n/2}\left({\textstyle\frac{b^2}{8a}}\right)\,e^{{\textstyle \frac{b^2}{8a}}}.
\end{equation}
Let us exploit this formula constructing the superposition of $\Psi_{mBG}^{(n)}$'s with additional Gaussian weighting factor $e^{-\chi^2/\kappa}$, i.e, defining 
\begin{equation}\label{supshbgmbg}
\Psi_s(r,\varphi,z)\stackrel{\mathrm{def}}{=}\int\limits_0^\infty d\chi\,e^{\textstyle{-\frac{\chi^2}{\kappa}}} \,\Psi_{mBG}^{(n)}(r,\varphi,z).
\end{equation}
where $\kappa$ satisfies the same convergence condition as in~(\ref{supgmbg}). Applying~(\ref{int4}) one gets
\begin{eqnarray}
\Psi_s(r,\varphi,z)=&&\sqrt{\frac{\pi\kappa}{\alpha(z)(\alpha(z)+\kappa)}}\,e^{in\varphi}\label{sshbgmbg}\\
&&\times e^{\textstyle -\frac{r^2}{2\alpha(z)}\,\frac{2\alpha(z)+\kappa}{\alpha(z)+\kappa}}I_{n/2}\left({\textstyle \frac{\kappa r^2}{2\alpha(z)(\alpha(z)+\kappa}}\right)\nonumber.
\end{eqnarray}
The halved index of the Bessel function is characteristic of a beam provisionally termed ``special hyperbolic Bessel-Gaussian beam'' in~\cite{trsh}. In order to transform this expression into the preferred form (i.e. formula (14) of~\cite{trsh} up to a normalization constant), it is sufficient to redefine the constants as follows (for real $\kappa$)
\begin{equation}\label{mobwah}
w_0^2\;\;\longmapsto \;\; w_0^2-\frac{\kappa}{2},\;\;\;\;\;\; \kappa\;\;\longmapsto \;\; 2\chi.
\end{equation}
The designation of the parameter $\chi$ occurring above is copied from the cited paper and has no relation to the integration variable of the formula~(\ref{supshbgmbg}).

In Appendix the expression~(\ref{sshbgmbg}) is derived again by directly solving the paraxial equation.

\subsection{Generalized (hyperbolic) paraxial beam}\label{genpmbg}

The following summation formula for the products of modified Bessel functions holds~\cite{watson}
\begin{eqnarray}
\sum_{n=-\infty}^\infty &&e^{i n\varphi}I_n(u)I_{m+n}(w)=\left(\frac{w+ue^{-i\varphi}}{w+ue^{i\varphi}}\right)^{m/2}\label{sumf}\\ &&\times I_m\left(\sqrt{w^2+u^2+2uw\cos\varphi}\right).\nonumber
\end{eqnarray}
It will be used below to derive the expansion of the ``generalized'' paraxial beam (that does not exhibit cylindrical character) onto cylindrical modified BG beams. Let us first substitute
\begin{equation}\label{df}
u=\mu,\;\;\;\;\;\; w=\frac{2\chi r}{\alpha(z)}, 
\end{equation}
where $\mu$ and $\chi$ are certain, possibly also complex constants, and create the superposition of mBG modes over the orbital angular momentum index $n$ with weights that in turn are modified Bessel functions of $\mu$. 

Using~(\ref{sumf}) one finds the needed sum 
\begin{eqnarray}
\sum_{n=-\infty}^\infty e^{i n\varphi}&&I_n(\mu)I_{m+n}\left({\textstyle\frac{2\chi r}{\alpha(z)}}\right)=e^{-im\varphi}\left(\frac{\xi(x,y,z)}{\eta(x,y,z)}\right)^m\nonumber\\ &&\times I_m\left(\xi(x,y,z)\eta(x,y,z)\right).\label{sumfa}
\end{eqnarray}
where 
\begin{subequations}\label{defvar}
\begin{align}
&\xi(x,y,z)=\left(\mu+\frac{2\chi}{\alpha(z)}(x+i y)\right)^{1/2},\label{defvarxi}\\
&\eta(x,y,z)=\left(\mu+\frac{2\chi}{\alpha(z)}(x-i y)\right)^{1/2}.\label{defvareta}
\end{align}
\end{subequations}
Now, multiplying both sides of~(\ref{sumfa}) by the factor 
\begin{equation}\label{fact}
\frac{1}{\alpha(z)}\, e^{{\textstyle - \frac{r^2+\chi^2}{\alpha(z)}}+i m\varphi},
\end{equation}
and shifting the dummy summation variable $n$ by $-m$ we come to the simple superposition formula spoken of above
\begin{eqnarray}
\Psi_s^{(m)}(x,y,z)=&&\frac{1}{\alpha(z)}\, e^{\textstyle - \frac{r^2+\chi^2}{\alpha(z)}}\left(\frac{\xi(x,y,z)}{\eta(x,y,z)}\right)^m\nonumber\\ &&\times I_m\left(\xi(x,y,z)\eta(x,y,z)\right)\label{supn6}\\=&&(-1)^m\sum_{n=-\infty}^\infty I_{n-m}(\mu)\Psi_{mBG}^{(n)}(r,\varphi,z).\nonumber
\end{eqnarray}
This expression represents a beam found in~\cite{trgen} by explicit solution of the paraxial equation and independently constructed from shifted Gaussian beams. Now its expansion onto mBG modes is found. It can be called ``generalized'' or ``few-parameter'' hyperbolic paraxial beam  since by a suitable choice of the parameters $\mu$ and $\chi$ some of the standard beams can be obtained. In particular the former is responsible for non-cylindrical shape of the intersection by some perpendicular plane. It becomes more and more axisymmetric when $\mu$ tends to zero (but paradoxically also when $\mu\rightarrow\infty$, since then in regions where the symmetry would be broken the wave is damped by the Gaussian factor).

\section{Superpositions of Bessel-Gaussian beams}\label{bgb}

Regular Bessel-Gaussian beams exhibit different character, which manifests itself through the existence of concentric rings of higher irradiance in the planes $z=\mathrm{const}$. Beams of the $n$th order are described with the following formula~\cite{gori}
\begin{equation}\label{bgbeam}
\Psi_{BG}^{(n)}(r,\varphi,z)=\frac{2}{\alpha(z)}e^{in\varphi}e^{{\textstyle \frac{\chi^2-r^2}{\alpha(z)}}}J_n\left({\textstyle\frac{2\chi r}{\alpha(z)}}\right),
\end{equation}
As before, two parameters can be used to build superpositions: the continuous parameter $\chi$ and the discrete one -- $n$.  Superpositions of this kind of modes with respect to the parameter $\chi$ are, however, somewhat more problematic than those for modified BG beams due to the positive sign in the exponent $e^{\chi^2/\alpha(z)}$ (one should remember that the real part of $\alpha(z)$ is always positive). This unwanted sign precludes all the weighting factors that exhibit the power-law behavior, i.e., without gaussians, that could potentially save the convergence of the infinite integral. Therefore, one has to inevitably introduce into the amplitudes the additional Gaussian factor in order to overcome the divergence at infinity. For this reason, it is not feasible to get $\gamma$ or Kummer-Gaussian beams from superpositions over the parameter $\chi$, where weighting factors would need to be purely power ones.

The BG beam (\ref{bgbeam}) can be constructed by superimposing slanted Gaussian beams~\cite{bagini}. In this case the parameter $\chi$, unlike Sec. \ref{supmbg}, is now associated with the semi-aperture angle of the constituent Gaussian beams and the Rayleigh length. It differs by the factor $z_R/k$~\cite{bor,mad} from the parameter $\beta$ of~\cite{gori} and~\cite{bagini}. Hence, there is no formal restriction on extending the superposition over $\chi$ to the entire positive semi-axis.

\subsection{Gaussian beam}\label{gaubg}

Gaussian beams can be constructed out of BG modes in the similar way as it was done with mBG modes, but with slight modifications. Instead of using the integral~(\ref{int1}), we now need~\cite{prudnikov}
\begin{equation}\label{int5}
\int\limits_0^\infty dx\, x^{n+1}e^{- a x^2} J_n(b x)= \frac{b^n}{(2a)^{n+1}}\,e^{{\textstyle \frac{b^2}{4a}}},
\end{equation}
where contrary to (\ref{ab}) one has to set
\begin{equation}\label{ab2}
a=\frac{1}{\kappa}-\frac{1}{\alpha(z)},\;\;\;\;\;\; b=\frac{2r}{\alpha(z)},
\end{equation} 

Let us now define the superposition of BG modes in the form
\begin{equation}\label{supgbg}
\Psi_s(r,\varphi,z)\stackrel{\mathrm{def}}{=}\int\limits_0^\infty d\chi\left(\frac{\chi}{\kappa}\right)^{n+1}\!\!\!\! e^{{\textstyle -\frac{\chi^2}{\kappa}}}\Psi_{BG}^{(n)}(r,\varphi,z).
\end{equation}
The new condition to be satisfied here is $\mathrm{Re}\,\kappa<w_0^2$, which ensures the convergence of the integral~(\ref{supgbg}) for arbitrary value of $z$. Substituting the explicit formula (\ref{bgbeam}) and exploiting the integral (\ref{supgbg}) one comes to
\begin{equation}\label{sgbg}
\Psi_s(r,\varphi,z)=\frac{1}{(\alpha(z)-\kappa)^{n+1}}r^ne^{in\varphi}e^{{\textstyle -\frac{r^2}{\alpha(z)-\kappa}}}
\end{equation}
Similarly to~(\ref{sgmbg}) this expression represents a standard Gaussian beam of the $n$th order with the obvious substitution
\begin{equation}\label{copa2}
\alpha(z)\;\;\longmapsto\;\; \alpha(z)-\kappa.
\end{equation}
which leads to the modification of the value of the waist $w_0\mapsto \sqrt{w_0^2-\mathrm{Re}\,\kappa}$ and to the shift of the focal plane along the $z$ axis, according to
\begin{equation}\label{mobwa}
w_0^2\;\;\longmapsto \;\; w_0^2-\mathrm{Re}\,\kappa,\;\;\;\;\;\; z\;\;\longmapsto \;\; z-\frac{1}{2}\,k\,\mathrm{Im}\,\kappa.
\end{equation}
Apparently it is the same result as in Sec.~\ref{gaumbg}, with merely inverted sign of the parameter $\kappa$. However, it should be kept in mind that modes with completely different properties are now being superposed, and also the parameter $\chi$ with respect to which the superposition proceeds has here a quite different physical sense than earlier. From this perspective, the almost identical result obtained from superpositions~(\ref{supgmbg}) and~(\ref{supgbg}) may seem somewhat puzzling. 
The above effect of the interference of BG beams is attributable to the tightening of the irradiance rings around the propagation axis with increasing value of $\chi$. Simultaneously, a large value of this parameter enforces shifting of the focus to the spot where the imaginary part of the coefficient of $\chi^2$ in the exponent, i.e., $1/\kappa-1/\alpha(z)$ tends to zero. Otherwise, the destructive interference occurs due to the rapid oscillations of the exponential function.

Let us now pass to the superposition of co-focal BG modes with respect to the topological indices $n$ and choosing certain constant value of $\chi$. The weighting factors are set, as in Sec.~\ref{gaumbg}, in the form of $e^{n v}$, dependent on some parameter $v$
\begin{equation}\label{supgbgn}
\Psi_s(r,\varphi,z)\stackrel{\mathrm{def}}{=}\sum_{-\infty}^\infty e^{nv}\Psi_{BG}^{(n)}(r,\varphi,z).
\end{equation}
The resulting beam can be rewritten in the explicit form as
\begin{equation}\label{supgnd1}
\Psi_s(r,\varphi,z)=\frac{2}{\alpha(z)}e^{{\textstyle \frac{\chi^2-r^2}{\alpha(z)}}}\sum_{-\infty}^\infty e^{n(v+i\varphi)}J_n\left({\textstyle\frac{2\chi r}{\alpha(z)}}\right).
\end{equation}
Now, using the standard formula for the generating function~\cite{abra}
\begin{equation}\label{generg}
e^{{\textstyle\frac{z}{2}}\left(t-{\textstyle\frac{1}{t}}\right)}=\sum_{n=-\infty}^\infty J_n(z)t^n
\end{equation}
we come to
\begin{eqnarray}
\Psi_s(r,\varphi,z)=&&\,\frac{2}{\alpha(z)}e^{{\textstyle\frac{\chi^2-r^2}{\alpha(z)}}}\label{supgnc1}\\
&&\times e^{{\textstyle\frac{2\chi r}{\alpha(z)}}\left(\cos\varphi\sinh v+i\sin\varphi\cosh v\right)}\nonumber\\
=&&\, \frac{2}{\alpha(z)} e^{{\textstyle-\frac{1}{\alpha(z)}}\,\left[(x-\chi\sinh v)^2+(y-i\chi\cosh v)^2\right]}\nonumber\\
=&&\, \Psi_{G\;\mathrm{sh}}^{(0)}(r,\varphi,z)\nonumber.
\end{eqnarray}
Thus, the only effect of combining BG modes instead of mBG ones (see Eq. (\ref{supgnc})) in the identical manner, is to obtain the shifted Gaussian beam by the slightly modified vector:
\begin{equation}\label{shvec}
\chi\left[\cosh v,i\sinh v\right]\;\;\longmapsto\;\; \chi\left[\sinh v,i\cosh v\right]
\end{equation}
Now, by picking a purely imaginary value of $v$, one also gets a purely imaginary displacement vector (in the case of superimposing mBG modes it became purely real).

The interference mechanism leading to shifted Gaussian beam is similar to that described in Sec.~\ref{gaumbg} with one essential difference. According to~(\ref{shvec}) for real $v$ the focus is shifted along the $x$-axis either to the right (for $v$ positive) or to the left (for $v$ negative). This was not the case for mBG modes since $\cosh v$ is always positive. This difference is attributable to the following property of the Bessel functions:
\begin{subequations}\label{probe}
\begin{align}
&J_{-n}(z)=(-1)^nJ_n(z),\label{probej}\\
&I_{-n}(z)=I_n(z)\label{probei}.
\end{align}
\end{subequations}
For BG modes and for positive $v$ the positive-$n$ modes contribution becomes more significant due to the increasing factors $e^{nv}$. Moreover the destructive interference occurs for the left half-plane (i.e. for $x<0$) -- due to the presence of $e^{in\varphi}$ -- and constructive one for the right half-plane ($x>0$). The irradiance peak moves, therefore to the right. When $v<0$ the situation is reversed. The negative-$n$ modes contribute more substantially and thanks to the negative sign in~(\ref{probej}) the destructive interference occurs now for $x>0$. This was not the case when considering the superposition of mBG modes, since the minus sign in~(\ref{probei}) is absent: the destructive interference takes always place for $x<0$ (i.e., for any $n$ and any $v$).

\subsection{Special hyperbolic Bessel-Gaussian beam}\label{spec1}

Consider now the integral similar to~(\ref{int4}) but involving regular Bessel function~\cite{prudnikov}:
\begin{equation}\label{int6}
\int\limits_0^\infty dx\, e^{-ax^2} J_n(b x)= \sqrt{\frac{\pi}{4a}}\,I_{n/2}\left({\textstyle\frac{b^2}{8a}}\right)\,e^{{\textstyle -\frac{b^2}{8a}}},
\end{equation}
and define the superposition 
\begin{equation}\label{supshbgbg}
\Psi_s(r,\varphi,z)\stackrel{\mathrm{def}}{=}\int\limits_0^\infty d\chi\,e^{\textstyle{-\frac{\chi^2}{\kappa}}} \,\Psi_{BG}^{(n)}(r,\varphi,z).
\end{equation}
where $\mathrm{Re}\,\kappa<w_0^2$, as in the previous section. After having performed the integration one gets the expression 
\begin{eqnarray}
\Psi_s(r,\varphi,z)=&&\sqrt{\frac{\pi\kappa}{\alpha(z)(\alpha(z)-\kappa)}}\,e^{in\varphi}\label{sshbgbg}\\
&&\times e^{\textstyle -\frac{r^2}{2\alpha(z)}\,\frac{2\alpha(z)-\kappa}{\alpha(z)-\kappa}}I_{n/2}\left({\textstyle \frac{\kappa r^2}{2\alpha(z)(\alpha(z)-\kappa}}\right)\nonumber.
\end{eqnarray}

If one now redefines the beam's waist and introduces some new parameter $\chi$ (as in~(\ref{mobwah}) this designation has nothing to do with the earlier integration variable) according to
\begin{equation}\label{mobwaf}
w_0^2\;\;\longmapsto \;\; w_0^2+\frac{\kappa}{2},\;\;\;\;\;\; \kappa\;\;\longmapsto \;\; 2\chi,
\end{equation}
once again the expression (14) of~\cite{trsh} is obtained.

It may seem somewhat puzzling that the identical superpositions of different beams (mBG and BG beams) yield the same result. It should be noted, however, that the BG and mBG beams are closely related to each other. In fact, they only differ by the replacement of the real integration parameter $\chi$ by the purely imaginary one. Since this is the case, it can be concluded that the formulas~(\ref{supshbgmbg}) and~(\ref{supshbgbg}) describe actually superpositions of the same modes, but one time along the real and the other along the imaginary axis of the complex $\chi$. These superpositions can produce the same result as long as the sign of $\kappa$ is inverted (and it is in fact the only distinction that is revealed in the formulas~(\ref{sshbgmbg}) and~(\ref{sshbgbg})) and the integrand function decreases sufficiently fast at infinity (relative to complex $\chi$). 

\subsection{Special Laguerre-Gaussian beam}\label{lagg}

The Laguerre-Gaussian beams are well known in the optical literature, both standard and elegant ones~\cite{sie,tan}. In this subsection it will be shown that there exists yet another kind of LG beams, which we provisionally call ``special'' LG beams (by their similarity to special hyperbolic BG beams), and which can be obtained by some specific superposition of BG modes. In order to derive the relevant formulas, let us use the integral~\cite{prudnikov}
\begin{eqnarray}
\int\limits_0^\infty dx\, x^{n+2p+1}&&\,e^{- a x^2} J_n(b x)\label{int7}=\\
&& \frac{p!\,b^n}{2^{n+1}a^{n+p+1}}\,e^{{\textstyle -\frac{b^2}{4a}}}L_p^n\left({\textstyle\frac{b^2}{4a}}\right),\nonumber
\end{eqnarray}
where $L_n^p(z)$ stands for generalized Laguerre polynomial, and consider the following superposition
\begin{equation}\label{supslgbg}
\Psi_s(r,\varphi,z)\stackrel{\mathrm{def}}{=}\int\limits_0^\infty d\chi\left(\frac{\chi}{\kappa}\right)^{n+2p+1}\!\!\!\! e^{{\textstyle -\frac{\chi^2}{\kappa}}}\Psi_{BG}^{(n)}(r,\varphi,z).
\end{equation}

Performing the integration according to~(\ref{int7}), one gets the following expression 
\begin{eqnarray}
\Psi_s(r,\varphi,z)=&&\frac{p!\, \alpha(z)^p}{(\alpha(z)-\kappa)^{n+p+1}}\, r^ne^{in\varphi}\label{sslgbg}\\
&&\times e^{\textstyle -\frac{r^2}{\alpha(z)-\kappa}}L_p^n\left({\textstyle \frac{\kappa r^2}{\alpha(z)(\alpha(z)-\kappa)}}\right).\nonumber
\end{eqnarray}
Up to our knowledge, this kind of a paraxial beam has not been considered so far in the literature and merits some separate detailed examination, which is, however, beyond the scope of the present study.

As it was in Sec~\ref{gaubg} and \ref{spec1} the convergence of the integral~(\ref{supslgbg}) requires that $\mathrm{Re}\,\kappa<w_0^2$. In the limit of vanishing $\kappa$, which is well defined (see remarks below formula~(\ref{supgmbg})), the Gaussian beam~(\ref{gauss}) is restored (apart from a constant factor), due to
\begin{equation}\label{limlg}
\lim_{z\to 0} L_p^n(z)=\left(\begin{array}{c} n+p \\ n\end{array}\right).
\end{equation} 
This result is understood if one remembers that any BG beam is composed of inclined Gaussian modes~\cite{bagini}. When $\kappa\rightarrow 0^+$, only that with vanishing value of $\chi$ (i.e. that with the wave vector parallel to the $z$ axis) contributes in the superposition~(\ref{supslgbg}) owing to the presence of $\exp({\textstyle -\frac{\chi^2}{\kappa}})$.

\subsection{Generalized paraxial beam}\label{genpbg}

The ``generalized'' or ``few-parameter'' paraxial beam which, in the hyperbolic version, was constructed in Sec.~\ref{genpmbg} out of mBG beams can also be obtained by the similar superposition of BG beams if the following summation formula~\cite{watson} is used:
\begin{eqnarray}
\sum_{n=-\infty}^\infty e^{i n\varphi}&&J_n(u)J_{m+n}(w)=\left(\frac{w-ue^{-i\varphi}}{w-ue^{i\varphi}}\right)^{m/2}\nonumber\\ &&\times J_m\left(\sqrt{w^2+u^2-2uw\cos\varphi}\right).\label{sumfb}
\end{eqnarray}
The values of the parameters $u$ and $w$ stay at our disposal, so let us choose them in a slightly different form then before:
\begin{equation}\label{df1}
u=\mu,\;\;\;\;\;\; w=-\frac{2\chi r}{\alpha(z)}, 
\end{equation}
where $\mu$ is a certain real or complex constant. Since $J_n(-z)=(-1)^nJ_n(z)$, this implies that
\begin{eqnarray}
\sum_{n=-\infty}^\infty&& (-1)^{n+m}e^{i n\varphi}J_n(\mu)J_{m+n}\left({\textstyle\frac{2\chi r}{\alpha(z)}}\right)=\label{sumfc}\\ &&e^{-im\varphi}\left(\frac{\xi(x,y,z)}{\eta(x,y,z)}\right)^m \!\! J_m\left(\xi(x,y,z)\eta(x,y,z)\right),\nonumber
\end{eqnarray}
where $\xi(x,y,z)$ and $\eta(x,y,z)$ were defined in (\ref{defvar}).  Multiplying now both sides of this relation by the factor 
\begin{equation}\label{facta}
\frac{1}{\alpha(z)}\, e^{{\textstyle - \frac{r^2-\chi^2}{\alpha(z)}}+i m\varphi},
\end{equation}
and replacing the summation index $n$ with $n+m$ we come to
\begin{eqnarray}
\Psi_s^{(m)}(x,y,z)=&&\frac{1}{\alpha(z)}\, e^{\textstyle - \frac{r^2-\chi^2}{\alpha(z)}}\left(\frac{\xi(x,y,z)}{\eta(x,y,z)}\right)^m\nonumber\\
&& \times J_m\left(\xi(x,y,z)\eta(x,y,z)\right)\label{supn8}\\
=&&\sum_{n=-\infty}^\infty (-1)^nJ_{n-m}(\mu)\Psi_{BG}^{(n)}(r,\varphi,z).\nonumber
\end{eqnarray}
The expression on the left hand side is a solution of the paraxial equation and constitutes the non-hyperbolic counterpart of the beam derived in~\cite{trgen}.

\section{Conclusions}\label{sum}

In this work, the results on various superpositions of Bessel-Gaussian beams and modified Bessel-Gaussian beams are presented in a fairly systematic way. The dependence of the relevant formulas on two parameters that are under our control, and with respect to which the superpositions can be realized, has been exploited. This seems to be the main benefit of using BG or mBG beams (versus, for example, LG beams): the presence of the extra parameter $\chi$ opens up the possibility of constructing with its use superpositions other than those created with respect to the discrete values of OAM. An analytical demonstration has been made of how, by selecting appropriate weighting functions for the constituent modes, a number of known beams, both cylindrical and asymmetric in nature, can be generated. 

It seems that the resulting analytical formulas contribute to a better unified description of different beams, and make the connection between various types of solutions to the paraxial equation~(\ref{paraxiala}) more apparent, which might also enhance the experimental possibilities of generating certain beams.

By superimposing BG modes, it has been possible to construct a beam that, to the best of our knowledge, has not appeared in the literature so far. It has been provisionally termed ``special Laguerre-Gaussian beam'' by analogy with previously introduced ``special hyperbolic Bessel-Gaussian beam''~\cite{trsh}. This new type of mode will be studied in detail elsewhere.
A resume of the expansions that have been possible to obtain in this work is provided in Table~\ref{ab}.

\begin{table}[h]
\setlength\extrarowheight{4pt}
\caption{Summary of the expansions}\label{ab}
\begin{ruledtabular}
\begin{tabular}{ccc}
beams & mBG & BG\\
\colrule
G  & $\chi$ integral (\ref{supgmbg}) & $\chi$ integral (\ref{supgbg})\\
 & sum over $n$(\ref{supgmbgn}) & sum over $n$ (\ref{supgbgn})\vspace{0.2em}\\
\colrule
$\gamma$ & $\chi$ integral (\ref{supgammbg})& \vspace{0.2em}\\
\colrule
KG & $\chi$ integral (\ref{supkgmbg}) & \vspace{0.2em}\\
\colrule
shBG & $\chi$ integral (\ref{supshbgmbg}) & $\chi$ integral (\ref{supshbgbg})\vspace{0.2em}\\
\colrule
sLG & & $\chi$ integral (\ref{supslgbg})\vspace{0.2em}\\
\colrule
Generalized & & sum over $n$ (\ref{supn8}) \vspace{0.2em}\\
\colrule
Generalized (hyp.) & sum over $n$ (\ref{supn6})& \vspace{0.2em}\\
\end{tabular}
\end{ruledtabular}
\end{table}

Additionally in the appendix the explicit derivation of the expression~(\ref{sshbgmbg}) describing a certain new asymmetric beam is demonstrated by directly solving the paraxial equation.

\appendix*
\section{Explicit derivation of the expression~(\ref{sshbgmbg}) from the paraxial equation}\label{app}

In this appendix the explicit formula~(\ref{sshbgmbg}) will be obtained (in the similar manner the formula (\ref{sshbgbg}) can be derived) by directly solving the paraxial equation, which in the cylindrical variables has the form 
\begin{equation}\label{paraxial}
\left(\partial_r^2+\frac{1}{r}\,\partial_r+\frac{1}{r^2}\,\partial_\varphi^2+2ik\partial_z\right)\Psi(r,\varphi,z)=0.
\end{equation}
Let us first separate the azimuthal dependence by setting
\begin{equation}\label{sub1}
\Psi(r,\varphi,z)\frac{1}{\sqrt{\alpha(z)}}\,e^{in\varphi}\tilde{\Psi}(r,z),
\end{equation}
which brings Eq.~\ref{paraxial} to the following form
\begin{equation}\label{paraxial1}
\left(\partial_r^2+\frac{1}{r}\,\partial_r-\frac{n^2}{r^2}+\frac{2}{\alpha(z)}+2ik\partial_z\right)\tilde{\Psi}(r,z)=0.
\end{equation}
Now in place of $r$ and $z$ we introduce two new variables:
\begin{equation}\label{tnv}
\xi=\frac{r^2}{\alpha(z)},\;\;\;\; \eta=\frac{1}{\alpha(z)+\kappa},
\end{equation}
where $\alpha(z)$ was defined in~(\ref{alpha}) and $\kappa$ is a certain constant. In these variables Eq.~(\ref{paraxial1}) can be rewritten as
\begin{eqnarray}
\bigg[\xi\partial_\xi^2+(1+\xi)\partial_\xi&-&\frac{(n/2)^2}{\xi}+\frac{1}{2}\label{paraxial2}\\
&+&\eta^2\left(\frac{1}{\eta}-\kappa\right)\partial_\eta\bigg]\tilde{\Psi}(\xi,\eta)=0.\nonumber
\end{eqnarray}
Let us now seek the function $\tilde{\Psi}(\xi,\eta)$ in the form
\begin{equation}\label{solu}
\tilde{\Psi}(\xi,\eta)=\sqrt{\eta}\,e^{-\xi}\Phi(\tau),
\end{equation}
where $\tau$ is a new variable defined with $\tau=\frac{1}{2}\,\kappa\xi\eta$. After having inserted~(\ref{solu}) into~(\ref{paraxial2}), it can be verified that following {\em ordinary} differential equation is obtained:
\begin{equation}\label{feq}
\tau\Phi''(\tau)+(1-2\tau)\Phi'(\tau)-\left(\frac{(n/2)^2}{\tau}+1\right)\Phi(\tau)=0,
\end{equation}
where primes denote the differentiations with respect to the variable $\tau$.
The final substitution
\begin{equation}\label{fis}
\Phi(\tau)=e^\tau f(\tau)
\end{equation}
leads to the modified Bessel equation:
\begin{equation}\label{mobe}
f''(\tau)+\frac{1}{\tau}\,f'(\tau)-\left(\frac{(n/2)^2}{\tau^2}
+1\right)f(\tau)=0,
\end{equation}
which has the general solution
\begin{equation}\label{geso}
f(\tau)=C_1I_{n/2}(\tau)+C_2K_{n/2}(\tau).
\end{equation}
Setting $C_2=0$ (the function $K_{n/2}(\tau)$ for $n>1$ does not lead to a normalizable solution in the perpendicular plane), one can write down the complete result in the form 
\begin{eqnarray}
\Psi(r,\varphi,z)=&&\,\frac{C_1}{\sqrt{\alpha(z)(\alpha(z)+\kappa)}}\,e^{in\varphi}e^{\textstyle-\frac{r^2}{\alpha(z)+\kappa}}\label{fsol}\\
&& \times e^{\textstyle -\frac{\kappa r^2}{2\alpha(z)(\alpha(z)+\kappa)}} I_{n/2}\left({\textstyle\frac{\kappa r^2}{2\alpha(z)(\alpha(z)+\kappa)}}\right)\nonumber\\
=&&\,C_1\sqrt{\frac{\pi}{\alpha(z)(\alpha(z)+\kappa)}}\,e^{in\varphi}\nonumber\\
&&\times e^{\textstyle -\frac{r^2}{2\alpha(z)}\,\frac{2\alpha(z)+\kappa}{\alpha(z)+\kappa}}I_{n/2}\left({\textstyle \frac{\kappa r^2}{2\alpha(z)(\alpha(z)+\kappa)}}\right)\nonumber,
\end{eqnarray}
to be compared with~(\ref{sshbgmbg}).

\end{document}